\documentstyle[12pt]{article}

\advance\textheight by \topskip
\input{epsf.sty}

\def\axowidth{0.5 }
\def\axoscale{1.0 }

\def\Gluon(#1,#2)(#3,#4)#5#6{
\put(0,0){
}
}
\def\Photon(#1,#2)(#3,#4)#5#6{
\put(0,0){
}
}
\def\PhotonArc(#1,#2)(#3,#4,#5)#6#7{
\put(0,0){
}
}
\def\GlueArc(#1,#2)(#3,#4,#5)#6#7{
\put(0,0){
}
}
\def\ArrowArc(#1,#2)(#3,#4,#5){
\put(0,0){
}
}
\def\LongArrowArc(#1,#2)(#3,#4,#5){
\put(0,0){
}
}
\def\DashArrowArc(#1,#2)(#3,#4,#5)#6{
\put(0,0){
}
}
\def\ArrowArcn(#1,#2)(#3,#4,#5){
\put(0,0){
}
}
\def\LongArrowArcn(#1,#2)(#3,#4,#5){
\put(0,0){
}
}
\def\DashArrowArcn(#1,#2)(#3,#4,#5)#6{
\put(0,0){
}
}
\def\ArrowLine(#1,#2)(#3,#4){
\put(0,0){
}
}
\def\LongArrow(#1,#2)(#3,#4){
\put(0,0){
}
}
\def\DashArrowLine(#1,#2)(#3,#4)#5{
\put(0,0){
}
}
\def\Line(#1,#2)(#3,#4){
\put(0,0){
}
}
\def\DashLine(#1,#2)(#3,#4)#5{
\put(0,0){
}
}
\def\CArc(#1,#2)(#3,#4,#5){
\put(0,0){
}
}
\def\DashCArc(#1,#2)(#3,#4,#5)#6{
\put(0,0){
}
}
\def\Vertex(#1,#2)#3{
\put(0,0){
}
}
\def\Text(#1,#2)[#3]#4{
\put(#1,#2){\makebox(0,0)[#3]{#4}}
}
\def\BCirc(#1,#2)#3{
\put(0,0){
}
}
\def\GCirc(#1,#2)#3#4{
\put(0,0){
}
}
\def\EBox(#1,#2)(#3,#4){
\put(0,0){
}
}
\def\BBox(#1,#2)(#3,#4){
\put(0,0){
}
}
\def\GBox(#1,#2)(#3,#4)#5{
\put(0,0){
}
}
\def\Boxc(#1,#2)(#3,#4){
\put(0,0){
}
}
\def\BBoxc(#1,#2)(#3,#4){
\put(0,0){
}
}
\def\GBoxc(#1,#2)(#3,#4)#5{
\put(0,0){
}
}

\def\fsize{10 }

\def\PText(#1,#2)(#3)[#4]#5{
\ifx#4 lt{\def\fmode{0 }}\else{
\ifx#4 tl{\def\fmode{0 }}\else{
\ifx#4 lb{\def\fmode{2 }}\else{
\ifx#4 bl{\def\fmode{2 }}\else{
\ifx#4 l{\def\fmode{1 }}\else{
\ifx#4 rt{\def\fmode{6 }}\else{
\ifx#4 tr{\def\fmode{6 }}\else{
\ifx#4 rb{\def\fmode{8 }}\else{
\ifx#4 br{\def\fmode{8 }}\else{
\ifx#4 r{\def\fmode{7 }}\else{
\ifx#4 t{\def\fmode{3 }}\else{
\ifx#4 b{\def\fmode{5 }}\else{ \def\fmode{4 } }\fi
}\fi}\fi}\fi}\fi}\fi}\fi}\fi}\fi}\fi}\fi}\fi
\put(#1,#2){\makebox(0,0)[]{\special{"
axodict begin
/pfont findfont /fsize scale setfont
0 0 #3 \fmode \fsize (#5) ptext
end }}}
}

\renewcommand{\baselinestretch}{1.2}
\renewcommand{\thefootnote}{\fnsymbol{footnote}}
\newcommand{\bea}{\begin{eqnarray}}
\newcommand{\eea}{\end{eqnarray}}
\renewcommand{\(}{\left(}
\renewcommand{\)}{\right)}
\renewcommand{\[}{\left[}
\renewcommand{\]}{\right]}
\newcommand{\sm}{Standard Model }

\newcounter{hilf}

\newcommand{\La}{\Lambda}
\newcommand{\Lambd}{{\it M}}

\begin{document}


\begin{titlepage}
  \renewcommand{\baselinestretch}{1}

  \thispagestyle{empty}
  
   {\bf \hfill                                       LMU--12/94
   } \\
  \vspace*{1.5cm}
  {\LARGE\bf
  \begin{center}         Exact Cancellation of
                         Quadratic \\ Divergences
                         in Top Condensation Models  \end{center}  }
  \vspace*{1cm}
  {\begin{center}        {\large Andreas Blumhofer}\footnote{Email:
                         ab at hep.physik.uni--muenchen.de}
   \end{center} }
  \vspace*{0cm}
  {\it \begin{center}    Sektion Physik                               \\
                         Ludwig--Maximilians--Universit\"at M\"unchen \\
                         Theresienstr.37, D--80333 M\"unchen
       \end{center} }
  \vspace*{2cm}
  {\Large \bf \begin{center} Abstract  \end{center}  }

  We discuss the hierarchy problem and the corresponding quadratic
  divergences in the top mode Standard Model. Quadratic divergences
  appear at each order $1/N_c$ since fermionic and bosonic contributions 
  are of different order $1/N_c$. It is shown that the full dynamical system 
  to all orders in $1/N_c$ admits a solution, where the sum of all quadratic 
  divergent contributions disappears.

  \renewcommand{\baselinestretch}{1.2}

\end{titlepage}

\newpage
\renewcommand{\thefootnote}{\arabic{footnote}}
\setcounter{footnote}{0}


\section{Introduction}

In spite of the successful description of the strong and electroweak
phenomena by the Standard Model, the prediction of the particle masses
is still a big challenge.
Although the Higgs sector of the \sm explains the electroweak
symmetry breaking and how particles acquire a mass, it only seems to be an
effective description due to the many parameters. We are still far 
away from a theory being able
to predict all known particle masses and mixing angles in the
Kobayashi--Maskawa--matrix. However in a first step such a theory should at 
least predict the W-- and Z--masses. If there is a great desert, then it should 
also explain, why the Grand Unification scale is so many orders of magnitude 
beyond the weak scale. In the \sm this leads to the so--called hierarchy 
problem, which manifests itself in quadratic divergences of the Higgs 
tadpoles and self--energies. Of course, if the \sm is the final theory, 
then quadratic divergences are eliminated by renormalization.
But if the divergences originate from a dynamical interaction
generating the particle masses, the cutoff $\La$ will be the scale of
that interaction. In that case the quadratic divergences give huge
contributions for the particle masses of the \sm which make their low
values unstable.

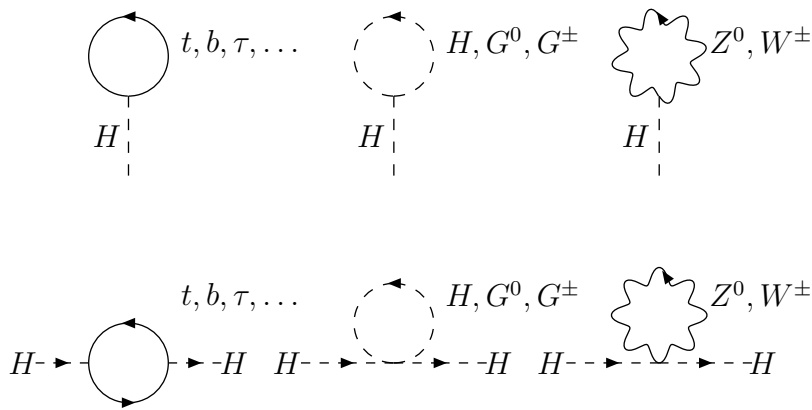
\begin{figure}[htb]
\begin{center}
\begin{picture}(290,80)(0,0)
\DashLine(45,10)(45,40)4
\ArrowArc(45,55)(15,270,269)
\Text(37,25)[c]{$H$}               \Text(65,60)[l]{$t,b,\tau,\ldots$}
\DashLine(145,10)(145,40)4
\DashArrowArc(145,55)(15,270,269)4
\Text(137,25)[c]{$H$}              \Text(165,60)[l]{$H,G^0,G^\pm$}
\DashLine(245,10)(245,40)4
\PhotonArc(245,55)(15,270,269)3 8
\ArrowLine(246,69)(245,70.5)
\Text(237,25)[c]{$H$}              \Text(265,60)[l]{$Z^0,W^\pm$}
\end{picture}
%
%
\begin{picture}(290,80)(0,0)
\DashArrowLine(10,20)(30,20)4
\DashArrowLine(60,20)(80,20)4
\ArrowArc(45,20)(15,0,180)         \ArrowArc(45,20)(15,180,360)
\Text(5,20)[c]{$H$}                \Text(85,20)[c]{$H$}
\Text(65,45)[l]{$t,b,\tau,\ldots$}
\DashArrowLine(110,20)(145,20)4
\DashArrowLine(145,20)(180,20)4
\DashArrowArc(145,35)(15,270,269)4
\Text(105,20)[c]{$H$}              \Text(185,20)[c]{$H$}
\Text(165,45)[l]{$H,G^0,G^\pm$}
\DashArrowLine(210,20)(245,20)4
\DashArrowLine(245,20)(280,20)4
\PhotonArc(245,38)(15,258.8,258.7)3 8
\ArrowLine(248,52)(247,53.5)
\Text(205,20)[c]{$H$}              \Text(285,20)[c]{$H$}
\Text(265,45)[l]{$Z^0,W^\pm$}
\end{picture}
\end{center}
\caption{\it One--loop quadratic divergent \sm graphs.}
\label{qudi}
\end{figure}

At the one--loop--level (fig.~\ref{qudi}) they have the form
\bea \[ 4\(m_t^2+m_b^2+\frac{1}{3}m_{\tau}^2+\ldots\)-M_H^2-M_Z^2-2M_W^2 \]
     \La^2,  \label{qd} \eea
%
%
where all fermions and bosons contribute according to their degrees of 
freedom. In trying to solve the hierarchy problem one can demand the
cancellation of quadratic divergences in lowest order,
namely the vanishing of (\ref{qd}), called Veltman condition,
first mentioned by R.~Decker and J.~Pestieau \cite{Veltman}. Unfortunately 
that condition requires an unnatural fine--tuning of independent masses, 
which is not enforced by symmetries. Moreover this relation is within the 
\sm completely regularization scheme dependent, i.~e. if we use
different cutoffs for different particle loops we get different conditions.
Hence in the pure \sm the Veltman condition has no meaning. This is different,
if the \sm arises as an effective low energy Lagrangian from some new
dynamics at a high scale. We will see that formfactors of composite 
particles and masses of the underlying theory act then like physical cutoffs
leading to an effective Veltman condition.

The regularization scheme dependence drops out, if a corresponding
symmetry is present. A big progress in this direction was made with 
supersymmetric field theories\footnote{for a review see \cite{Nilles}},
which are free of quadratic divergences because of the same bosonic and 
fermionic degrees of freedom of the superpartners. So far supersymmetry 
is the only theory which avoids that hierarchy problem.
Nevertheless the many new parameters in supersymmetric theories make the
particle mass problem worse since the Higgs mechanism is not replaced
by an underlying theory but is legitimated as fundamental.
Thus many authors tried to solve the hierarchy problem for the
\sm and for non--supersymmetric extensions \cite{Authors,Vargas,Jack}.
They dealt with the Veltman condition and further additional assumptions.
The two loop cancellation of quadratic divergences in the \sm \cite{Vargas} 
leads however to a further condition and higher orders surely make the 
situation even worse.
Some authors emphasized that in renormalizable scalar theories and sigma
models the higher order quadratic divergences are related to the
lowest order ones \cite{Jack}. But this theorem failed at the
four--loop--level.

It seems that a cancellation does not happen by accident. In the spirit of
supersymmetry one needs an additional relation between fermions and bosons 
leading to cancellations in a natural way. Nambu first argued that in a 
composite Higgs model the masses of the expression (\ref{qd}) are not 
independent and could automatically be arranged in such a way that the 
quadratic divergences cancel \cite{Nambu}.
In the proposed model without new fermions the Higgs mainly consists of a
top--antitop--pair tightly bounded by new interactions.
One expects therefore $M_H\approx 2m_t$, which roughly coincides with the
vanishing of (\ref{qd}).
Nambu introduced the term ``Quasi--Supersymmetry'' because the Higgs and the
top play the role of ``Superpartners'' in this model.
But in such theories the Higgs is a dynamical object and all masses
result from a complicated system of Schwinger--Dyson--equations.
Since dynamics and not symmetry sets the mass relations it is however hard 
to believe that they fulfill the Veltman condition exactly.

In a series of papers \cite{Bardeen} the top condensation idea was worked out
by several authors. Nevertheless the meaning of the Veltman condition for
top condensation was so far unclear.
In a recent paper \cite{B} it was shown that a cancellation of
quadratic divergences appears in such top condensation models in lowest
order of a combined color--flavor--expansion. The reason is that like
in supersymmetry the fermion and boson couplings, i.~e. in top condensation
the top Yukawa--coupling and the Higgs self--coupling, are related since the 
Higgs is a $\bar{t}t$--boundstate. In addition the number of fermion and boson
degrees of freedom are not equal in contrast to supersymmetry but matched in
such a way that the quadratic divergences cancel. This cancellation does 
however not require a mass relation between the top and the
Higgs mass and the Veltman condition, which was the motivation for this
scenario, disappears.
Nevertheless so far no symmetry was found as the origin of the cancellation
and the behaviour of the higher order $1/N_c$--corrections was therefore 
unclear.

Now in this paper I want to emphasize that the above mechanism seems to 
work, if we include all orders $1/N_c$, so that the quadratic divergences
cancel exactly. In section II top condensation and how the hierarchy 
problem appears in that model is recapitulated.
Section III explains the mechanism leading to the cancellation of
quadratic divergences in top condensation for a certain approximation
independent of the values of top and Higgs mass. That cancellation has 
another meaning than the usual Veltman condition. Contrasts and relations
are shown in section IV.
The rather technical section V includes a complete proof of the
cancellation to all orders, which follows from a solution of the full
system of Schwinger--Dyson--equations.
Section VI emphasizes that in a general context this solution does not appear
for other gauge groups
and prefers the \sm as the only possible choice of group parameters
leading to a cancellation of quadratic divergences.
Finally section VII handles with the vector--boson contributions and their
influence on the cancellation condition.


\section{The hierarchy problem in top condensation}

Let us first consider the minimal top condensation model in lowest order
$1/N_c$. It consists of the \sm without Higgs sector but with the additional 
four--fermion--interaction
\bea  {\cal L}_I =
    \frac{G}{N_c}\Big(\overline{\psi_L}t_R\Big)\Big(\overline{t_R}\psi_L\Big)
    \;\;;\;\;\;\; \psi_L = {t_L \choose b_L}\;\;, \label{l} \eea
where color is summed in the brackets. This model was studied by W.~A.~Bardeen, 
C.~T.~Hill and M.~Lindner (BHL) \cite{Bardeen}. 
We imagine that this four--fermion--interaction is generated by
new heavy bosons at a high scale $\Lambd$, far above the TeV--region.
There\-fore higher dimensional operators  are suppressed and the special
structure of this new interaction drops out \cite{Hasenfratz}.

After a reasonable choice of the auxiliary fields $H$, $G^0$ and $G^\pm$
one finds the usual \sm Higgs sector, where the kinetic terms for the
Higgs field components are absent:
\bea   {\cal L}_I &=& -\frac{1}{2} \frac{g_t^2N_c}{G}
     \(HH+G^0G^0+2G^+G^-\) \nonumber \\ &&
     -\frac{g_t}{\sqrt{2}}H\bar{t}t+i\frac{g_t}{\sqrt{2}}G^0\bar{t}\gamma^5t
     +g_t\(G^+\bar{t}Lb+G^-\bar{b}Rt\).   \label{tcl}
\eea
At tree level the Higgs and the Goldstone bosons do not propagate. Their
``static'' propagator has the form:
\bea i\frac{G}{g_t^2 N_c}. \eea
Two Yukawa couplings connected by that static propagator is nothing else than
a four--fermion--vertex. The Lagrangian (\ref{tcl}) is therefore completely
equivalent to the four--fermion--interaction (\ref{l}).

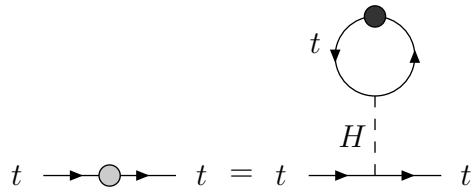
\begin{figure}[htb]
\begin{center}
\begin{picture}(190,80)(0,0)
\ArrowLine(20,10)(45,10)        \ArrowLine(45,10)(70,10)
\GCirc(45,10)4 {0.8}
\Text(10,10)[c]{$t$}            \Text(80,10)[c]{$t$}
\Text(95,10)[c]{=}
\ArrowLine(120,10)(145,10)      \ArrowLine(145,10)(170,10)
\DashLine(145,10)(145,40)4
\ArrowArc(145,55)(15,270,90)    \ArrowArc(145,55)(15,90,270)
\GCirc(145,70)4 {0.2}
\Text(110,10)[c]{$t$}             \Text(180,10)[c]{$t$}
\Text(137,25)[c]{$H$}             \Text(123,60)[c]{$t$}
\end{picture}
\end{center}
\caption{\it Gap equation in lowest order $1/N_c$.}
\label{gap}
\end{figure}

The top mass is generated by the gap equation which reads in lowest order
$1/N_c$ (see fig.~\ref{gap}):
\bea   1=\frac{2G}{(4\pi)^2}\(\Lambd^2-m_t^2\ln\frac{\Lambd^2}{m_t^2}\).
       \label{ge}
\eea
%
%
The bright circle is the self--energy and the dark circle the propagator, 
where the self--energies are summed up.
%
\begin{figure}[htb]
\begin{center}
\begin{picture}(270,50)(0,0)
\DashArrowLine(20,20)(50,20)4   \DashArrowLine(80,20)(110,20)4
\ArrowArc(65,20)(15,0,180)      \ArrowArc(65,20)(15,180,360)
\GCirc(65,35)4 {0.2}            \GCirc(65,5)4 {0.2}
\Text(10,26)[c]{$H$}            \Text(120,26)[c]{$H$}
\Text(10,14)[c]{$G^0$}          \Text(120,14)[c]{$G^0$}
\Text(65,47)[c]{$t$}            \Text(65,-7)[c]{$t$}
\DashArrowLine(160,20)(190,20)4 \DashArrowLine(220,20)(250,20)4
\ArrowArc(205,20)(15,0,180)     \ArrowArc(205,20)(15,180,360)
\GCirc(205,35)4 {0.2}
\Text(150,20)[c]{$G^-$}         \Text(260,20)[c]{$G^-$}
\Text(205,47)[c]{$t$}            \Text(205,-7)[c]{$b$}
\end{picture}
\end{center}
\caption{\it Boson self--energies in lowest order $1/N_c$.}
\label{boson}
\end{figure}
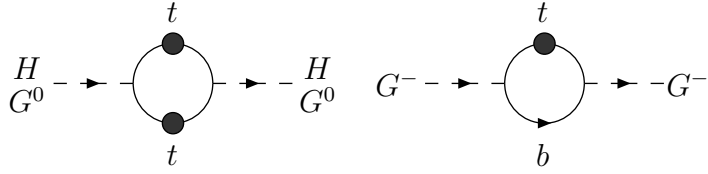
%

Including the resummation of the lowest order $1/N_c$ boson graphs
(fig.~\ref{boson})
the auxiliary fields become propagating boundstates. The propagators are:
\bea   D_H &=& \zeta^{-1}\frac{i}{p^2-4m_t^2}  \label{hp} \\
       D_{G^0} &=& \zeta^{-1}\frac{i}{p^2}  \\
       D_{G^\pm} &=& \zeta^{-1}\frac{i}{p^2}
\eea
where
\bea \zeta = \frac{N_cg_t^2}{(4\pi)^2}\ln\frac{\Lambd^2}{p^2}\label{ze} \eea
up to finite terms. From (\ref{hp}) we get the well known relation $M_H=2m_t$.

\begin{figure}[htb]
\begin{center}
\begin{picture}(270,50)(0,0)
\Photon(20,20)(50,20)2 3        \Photon(80,20)(110,20)2 3
\ArrowArc(65,20)(15,0,180)      \ArrowArc(65,20)(15,180,360)
\GCirc(65,35)4 {0.2}            \GCirc(65,5)4 {0.2}
\Text(10,20)[c]{$Z^0$}          \Text(120,20)[c]{$Z^0$}
\Text(65,47)[c]{$t$}            \Text(65,-7)[c]{$t$}
\Photon(160,20)(190,20)2 3      \Photon(220,20)(250,20)2 3
\ArrowArc(205,20)(15,0,180)     \ArrowArc(205,20)(15,180,360)
\GCirc(205,35)4 {0.2}           
\Text(150,20)[c]{$W^-$}         \Text(260,20)[c]{$W^-$}
\Text(205,47)[c]{$t$}           \Text(205,-7)[c]{$b$}
\end{picture}
\end{center}
\caption{\it Vector--boson self--energies in lowest order $1/N_c$.}
\label{vectorboson}
\end{figure}
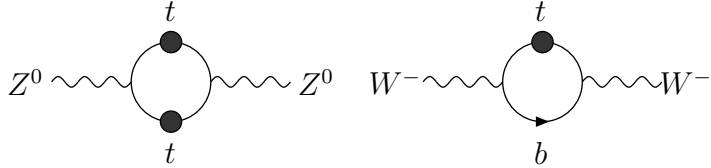

An important test for a mechanism explaining the electroweak symmetry
breaking is the right prediction of the W-- and Z--mass. The main
contributions are the fermion--loop diagrams in fig.~\ref{vectorboson}.
%
%
In the minimal model we find:
\bea M_W^2 &=& \frac{N_c g_2^2}{2(4\pi)^2}m_t^2\ln\frac{\Lambd^2}{m_t^2} \\
     M_Z^2 &=& \frac{N_c \(g_1^2+g_2^2\)}{2(4\pi)^2}
               m_t^2\ln\frac{\Lambd^2}{m_t^2}
\eea
which predicts the $M_Z/M_W$ ratio of the \sm and leads to the condition:
\bea 1=\frac{N_cg_t^2}{(4\pi)^2}\ln\frac{\Lambd^2}{m_t^2}. \eea
For $g_t\approx 1$ one gets $\Lambd\approx 5\cdot 10^{13}GeV$.
For such a high cutoff the gap equation (\ref{ge}) runs into a fine--tuning
problem because $m_t\approx\Lambd$ for almost the whole range of the
G--parameter.

Therefore more realistic top condensation models \cite{Hill} dealt with a
TeV--cutoff, where the predicted ratio for $M_W/m_t$ is to low.
These models include a strong interaction with new heavy gauge bosons
with mass $M_X$, which reproduces the four--fermion--interaction (\ref{l})
in the low energy limit.
%
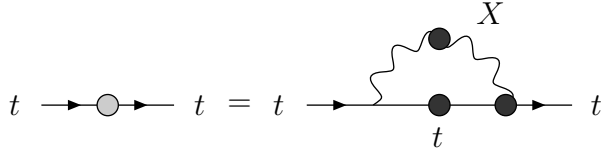
\begin{figure}[htb]
\begin{center}
\begin{picture}(240,80)(0,0)
\ArrowLine(20,10)(45,10)        \ArrowLine(45,10)(70,10)
\GCirc(45,10)4 {0.8}
\Text(10,10)[c]{$t$}            \Text(80,10)[c]{$t$}
\Text(95,10)[c]{=}
\ArrowLine(120,10)(145,10)      \ArrowLine(145,10)(195,10)
\ArrowLine(195,10)(220,10)
\PhotonArc(170,10)(25,0,180)3 6
\GCirc(170,10)4 {0.2}
\GCirc(195,10)4 {0.2}           \GCirc(170,35)4 {0.2}         
\Text(110,10)[c]{$t$}           \Text(230,10)[c]{$t$}
\Text(190,45)[c]{$X$}           \Text(170,-2)[c]{$t$}
\end{picture}
\end{center}
\caption{\it Gap equation with massive vector--bosons.}
\label{MX}
\end{figure}
%
The corresponding gap equation (fig.~\ref{MX}) provides a $p^2/M_X^2$ growth
of the dynamical top mass below $M_X$. This is closely related to the 
appearence of quadratic divergences in the low energy regime where only a
four--fermion--interaction remains. For large $M_X$ this strong running
provides the hierarchy problem, i.~e. the gap equation can not stabilize a low
top mass apart from a very tiny window of couplings according to
eq.~(\ref{ge}).

The situation drastically changes, if the low energy theory is free of
quadratic divergences which enables the top mass to run logarithmically
towards small $p^2$ after shortly falling down at the scale $M_X$.
This coincides with the usual renormalization group equation of the \sm where
a large enough top mass is running very smoothly up to the Landau pole, where
it blows up very fast.

In the following we will show that the low energy theory is indeed free of
quadratic divergences, so that the scale $M_X$ can be chosen very high to
get a reasonable W--mass value without running into a fine--tuning problem.


\section{Cancellation of quadratic divergences in a preliminary approximation}

So far the gap equation (\ref{ge}) is quadratic divergent. After including of
higher order diagrams the quadratic divergent contributions need a reshuffling
of Feynman graphs to be obvious. We will further show in the following the
cancellation of these contributions in a preliminary approximation.

From eq.~(\ref{hp}) we have seen that the lowest order $1/N_c$ Higgs 
propagator yields the relation $M_H=2m_t$. We therefore expect that a 
cancellation of quadratic divergences appears, at least partly, due to 
the Veltman condition.
Unfortunately this is not the case, since the lowest order $1/N_c$ gap
equation (fig.~\ref{gap}) contains only a fermion--loop tadpole. All the
other tadpoles of fig.~\ref{qudi} with the opposite sign are of higher
order $1/N_c$ or $g_{1/2}^2$. They drop out although Higgs and Goldstone
boson diagrams are numerically of the same order as the fermion tadpole.
Hence in lowest order $1/N_c$ it is impossible to see any cancellation.
From that one has to go beyond the leading order.
We neglect at the moment vector--boson contributions which will be considered
later.

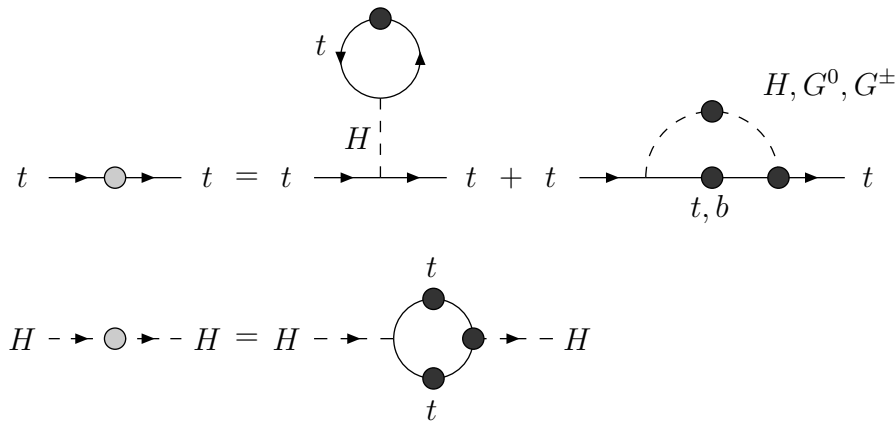
\begin{figure}[htb]
\begin{center}
\begin{picture}(310,80)(0,0)
\ArrowLine(20,10)(45,10)        \ArrowLine(45,10)(70,10)
\GCirc(45,10)4 {0.8}
\Text(10,10)[c]{$t$}            \Text(80,10)[c]{$t$}
\Text(95,10)[c]{=}
\ArrowLine(120,10)(145,10)      \ArrowLine(145,10)(170,10)
\DashLine(145,10)(145,40)4
\ArrowArc(145,55)(15,270,90)    \ArrowArc(145,55)(15,90,270)
\GCirc(145,70)4 {0.2}
\Text(110,10)[c]{$t$}             \Text(180,10)[c]{$t$}
\Text(137,25)[c]{$H$}             \Text(123,60)[c]{$t$}
\Text(195,10)[c]{+}
\ArrowLine(220,10)(245,10)      \ArrowLine(245,10)(295,10)
\ArrowLine(295,10)(320,10)
\DashCArc(270,10)(25,0,180)4
\GCirc(270,10)4 {0.2}
\GCirc(295,10)4 {0.2}           \GCirc(270,35)4 {0.2}
\Text(210,10)[c]{$t$}           \Text(330,10)[c]{$t$}
\Text(290,45)[l]{$H,G^0,G^\pm$} \Text(270,-2)[c]{$t,b$}
\end{picture}
\begin{picture}(310,60)(0,0)
\DashArrowLine(20,10)(45,10)4   \DashArrowLine(45,10)(70,10)4
\GCirc(45,10)4 {0.8}
\Text(10,10)[c]{$H$}            \Text(80,10)[c]{$H$}
\Text(95,10)[c]{=}
\DashArrowLine(120,10)(150,10)4 \DashArrowLine(180,10)(210,10)4
\ArrowArc(165,10)(15,0,180)     \ArrowArc(165,10)(15,180,360)
\GCirc(165,25)4 {0.2}           \GCirc(165,-5)4 {0.2}
\GCirc(180,10)4 {0.2}
\Text(165,37)[c]{$t$}         \Text(165,-17)[c]{$t$}
\Text(110,10)[c]{$H$}           \Text(220,10)[c]{$H$}
\end{picture}
\end{center}
\caption{\it Exact Schwinger--Dyson--equations.}
\label{exsd}
\end{figure}

\begin{figure}[htb]
\begin{center}
\begin{picture}(400,80)(0,0)
\ArrowLine(20,10)(45,10)        \ArrowLine(45,10)(70,10)
\GBoxc(45,10)(8,8){0.8}
\Text(10,10)[c]{$t$}            \Text(80,10)[c]{$t$}
\Text(95,10)[c]{:=}
\ArrowLine(120,10)(145,10)      \ArrowLine(145,10)(170,10)
\DashLine(145,10)(145,40)4
\ArrowArc(145,55)(15,270,90)    \ArrowArc(145,55)(15,90,270)
\GCirc(145,70)4 {0.2}
\Text(110,10)[c]{$t$}           \Text(180,10)[c]{$t$}
\Text(137,25)[c]{$H$}           \Text(123,60)[c]{$t$}
\Text(195,10)[c]{=}
\ArrowLine(220,10)(245,10)      \ArrowLine(245,10)(270,10)
\DashLine(245,10)(245,40)4
\ArrowArc(245,55)(15,270,90)    \ArrowArc(245,55)(15,90,270)
\GBoxc(245,70)(8,8) {0.2}
\Text(210,10)[c]{$t$}           \Text(280,10)[c]{$t$}
\Text(237,25)[c]{$H$}           \Text(223,60)[c]{$t$}
\Text(295,10)[c]{+}
\ArrowLine(320,10)(345,10)      \ArrowLine(345,10)(370,10)
\DashLine(345,10)(345,40)4
\CArc(345,55)(15,270,90)        \CArc(345,55)(15,90,270)
\DashArrowArc(345,70)(15,-30,210)4 \GCirc(345,85)4 {0.2}
\GBoxc(345,70)(8,8) {0.2}
\GBoxc(359,50)(8,8) {0.2}       \GBoxc(331,50)(8,8) {0.2}
\Text(310,10)[c]{$t$}           \Text(380,10)[c]{$t$}
\Text(337,25)[c]{$H$}           \Text(323,60)[c]{$t$}
\Text(395,10)[l]{$+\ldots$}
\end{picture}
\end{center}
\caption{\it Redefined top self--energy.}
\label{red}
\end{figure}
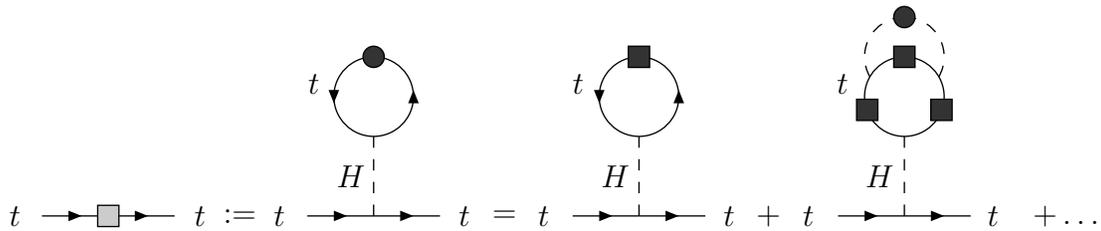

The exact Schwinger--Dyson--equations for the top and Higgs self--energies
are shown in fig.~\ref{exsd}.
%
%
The full vertices are defined by their skeleton expansion. The Higgs
line in the tadpole must be the non--propagating auxiliary field with
bare vertices to avoid double counting\footnote{E.g. if one inserts a
fermion--loop in the Higgs propagator, the upper part of the diagram is a
top self--energy correction for the tadpole with the new top--loop. This graph
is already considered in fig.~\ref{exsd}.}.
Tadpoles with boson--loops as in fig.~\ref{qudi} again do not appear in the
top self--energy. They would provide quadratic divergences with opposite
sign to cancel the fermion contribution. These graphs are implicitly
involved by reinsertion of the last diagram into the tadpole.
To make them obvious one has to extract the tadpole contribution as in the
recent paper \cite{B}. We define the tadpole by the box (see fig.~\ref{red})
and replace all fermion self--energy circles with boxes according to the
first line in fig.~\ref{exsd}, i.e. the last diagram of the first line in
fig.~\ref{exsd} is not included in the box and must be explicitly inserted in
the fermion--loop of the tadpole several times, where the full vertex is
replaced by the skeleton expansion.

%
%

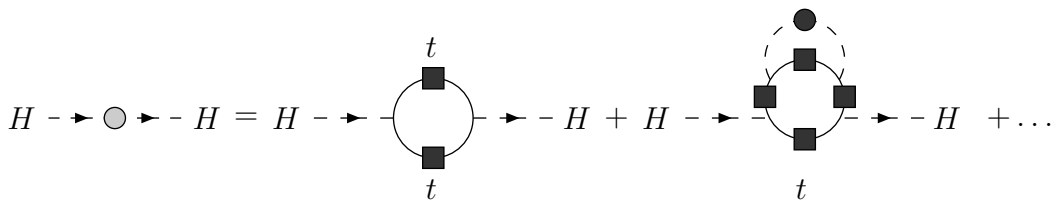
\begin{figure}[htb]
\begin{center}
\begin{picture}(420,60)(0,0)
\DashArrowLine(20,10)(45,10)4   \DashArrowLine(45,10)(70,10)4
\GCirc(45,10)4 {0.8}
\Text(10,10)[c]{$H$}            \Text(80,10)[c]{$H$}
\Text(95,10)[c]{=}
\DashArrowLine(120,10)(150,10)4 \DashArrowLine(180,10)(210,10)4
\ArrowArc(165,10)(15,0,180)     \ArrowArc(165,10)(15,180,360)
\GBoxc(165,25)(8,8) {0.2}       \GBoxc(165,-5)(8,8) {0.2}
\Text(165,37)[c]{$t$}         \Text(165,-17)[c]{$t$}
\Text(110,10)[c]{$H$}           \Text(220,10)[c]{$H$}
\Text(235,10)[c]{+}
\DashArrowLine(260,10)(290,10)4 \DashArrowLine(320,10)(350,10)4
\ArrowArc(305,17)(15,0,180)     \ArrowArc(305,17)(15,180,360)
\GBoxc(305,32)(8,8) {0.2}       \GBoxc(305,2)(8,8) {0.2}
\DashArrowArc(305,32)(15,-30,210)4 \GCirc(305,47)4 {0.2}
\GBoxc(290,18)(8,8) {0.2}       \GBoxc(320,18)(8,8) {0.2}
\Text(305,-17)[c]{$t$}
\Text(250,10)[c]{$H$}           \Text(360,10)[c]{$H$}
\Text(375,10)[l]{$+\ldots$}
\end{picture}
\end{center}
\caption{\it Exact Higgs boson self--energy.}
\label{reb}
\end{figure}

Since the tadpole just provides a mass correction for the top quark, this
also insures that the top propagator with the box does not receive wave
function corrections, which is essential for further discussions.
Using this new definition the boson self--energies can also be written in 
terms of the box propagator (see fig.~\ref{reb}).

%
%

In fig.~\ref{red} and fig.~\ref{reb} we get diagrams which appear
if we replace the three-- and four--boson--vertices of the \sm
in fig.~\ref{qudi} by fermion--loops.
Such fermion--loop induced vertices can easily be calculated up to finite
terms. One finds a simple relation between the fermion--loop induced and
the corresponding \sm tree vertices.
One must only multiply the \sm vertex $\Gamma_{\rm SM}$ with the following
factor to get the top condensation effective vertex $\Gamma$:
\bea  \Gamma = \Gamma_{\rm SM}\, \zeta \(\frac{2m_t}{M_H}\)^2+
      \mbox{finite terms}.  \label{vertex} \eea
The $\zeta$ is defined in eq.~(\ref{ze}) and depends on the largest momentum
in the diagram.
This factor is universal for all Higgs and Goldstone boson vertices.
All higher n--point functions, which do not exist in the Standard Model, are
finite in top condensation and therefore negligible for the discussion of the
leading divergences.

In the following we discuss the quadratic divergences of the Higgs 
self--energy. For simplicity we consider at the moment only self--energies
with at most one inner boson line.
To avoid an overkill of dots, boxes, etc. we do not write the box for
the fermion propagator and the dot for the boson propagator explicitly
from now on.

\begin{figure}[htb]
\begin{center}
\begin{picture}(250,90)(-30,10)
\DashArrowLine(-20,20)(10,20)4       \DashArrowLine(60,20)(90,20)4
\ArrowArc(35,20)(25,0,180)           \ArrowArc(35,20)(25,180,360)
\Text(-25,20)[c]{$H$}                \Text(95,20)[c]{$H$}
\Text(60,45)[l]{$t$}
\DashArrowLine(125,20)(155,20)4      \DashArrowLine(185,20)(215,20)4
\DashArrowArc(170,55)(25,270,269)4
\GCirc(170,27){15} 1
\ArrowArc(170,27)(15,45,135)         \ArrowArc(170,27)(15,135,225)
\ArrowArc(170,27)(15,225,315)        \ArrowArc(170,27)(15,315,45)
\Text(120,20)[c]{$H$}                \Text(220,20)[c]{$H$}
\Text(200,55)[l]{$H,G^0,G^\pm$}
\end{picture}
\end{center}
\caption{\it Quadratic divergent graphs in a preliminary approximation.}
\label{prap}
\end{figure}
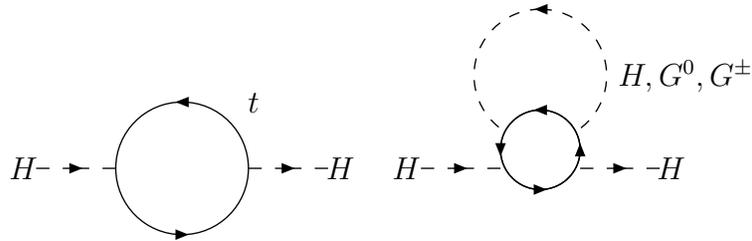

In our preliminary approximation we consider the remaining graphs in
fig.~\ref{prap}, where the four--boson--vertex is simply replaced by a fermion
loop\footnote{In the second diagram all topologies of the fermion--loop 
generated by the order of passing through the diagram are
involved although only one is shown in fig.~\ref{prap}.}.
%
%
To calculate the quadratic divergences one can use the \sm formulae for
both graphs, where in top condensation two additional factors enter the
formula for the second
graph. The propagator gets an $\zeta^{-1}$ due to eq.~(\ref{hp}), the 
effective vertex a factor $\zeta (2m_t/M_H)^2$ from 
eq.~(\ref{vertex})\footnote{Both $\zeta$'s depend on the same loop momentum 
and cancel therefore.}.
We end up with the top condensation expression:
\bea   \[ 4m_t^2-M_H^2\cdot\zeta^{-1}\cdot\zeta\(\frac{2m_t}{M_H}\)^2\]
       \Lambd^2 \label{prca} \eea
which surprisingly vanishes independent of the values of the top and the
Higgs mass. The reason for the cancellation is that the effective
four--boson--vertex and the top Yukawa--coupling are related in a certain way
and that the degrees of freedom are matched.
Nevertheless the cancellation happens for a mixed order $1/N_c$.
The first graph is of the order $N_c$ the second of the order
1\footnote{Each fermion--loop provides a factor $N_c$ and each boson line a
factor $1/N_c$.}.
The expression (\ref{prca}) has the following general form:
\bea   \[ \frac{4N_c}{3} m_t^2-M_H^2\cdot\zeta^{-1}\cdot\zeta
          \(\frac{2m_t}{M_H}\)^2\] \Lambd^2,  \eea
which only vanishes  for $N_c=3$. The cancellation obviously depends on the 
group structure. On the contrary the $M_H\!=\!2m_t$ relation
from eq.~(\ref{hp}) does not depend on $N_c$. Thus the similarity of both
equations is non--trivial.

Although the graphs in fig.~\ref{prap} are of different order $1/N_c$, 
both diagrams are of the same order of magnitude. Instead of the color
factor $N_c$ the second graph is enhanced by the number of the Higgs field
components.
Hence, in a recent paper \cite{B}, I have used a combined expansion in the
color and flavor degrees, where both graphs are of the same order.
In this paper we do not need any special expansion since we will calculate
the quadratic divergences to all orders $1/N_c$.

%
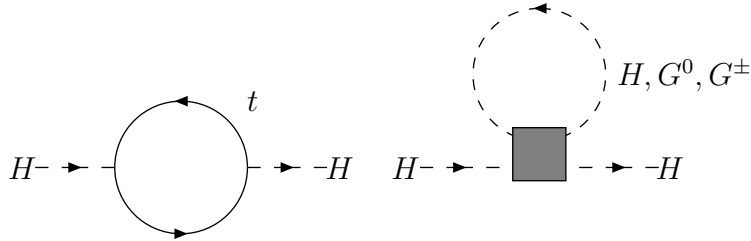
\begin{figure}[htb]
\begin{center}
\begin{picture}(250,90)(-30,10)
\DashArrowLine(-20,20)(10,20)4       \DashArrowLine(60,20)(90,20)4
\ArrowArc(35,20)(25,0,180)           \ArrowArc(35,20)(25,180,360)
\Text(-25,20)[c]{$H$}                \Text(95,20)[c]{$H$}
\Text(60,45)[l]{$t$}
\DashArrowLine(125,20)(155,20)4      \DashArrowLine(185,20)(215,20)4
\DashArrowArc(170,55)(25,270,269)4
\GBox(160,15)(180,35){0.5}
\Text(120,20)[c]{$H$}                \Text(220,20)[c]{$H$}
\Text(200,55)[l]{$H,G^0,G^\pm$}
\end{picture}
\end{center}
\caption{\it All order quadratic divergent graphs.}
\label{aoqd}
\end{figure}
%
We get the full set of the Higgs self--energy diagrams in fig.~\ref{reb} by 
replacing the fermion--loop in fig.~\ref{prap} by the full four--boson--vertex, 
the box in fig.~\ref{aoqd}.
It is clear that the quadratic divergences cancel to all orders if the
full vertex is just the fermion--loop induced vertex. This happens indeed, 
which we will see in section V.


\section{Low energy limit versus effective theory}

Before we go into technical details of the full analysis I like to
clarify the meaning of the cutoff $\Lambd$ for the cancellation.
Since the Veltman condition is regularization scheme dependent one has to
worry about the use of equal cutoffs in eq.~(\ref{prca}).
In contrast to the \sm we must realize that the four--boson--vertex is
dynamically generated and that both graphs of fig.~\ref{aoqd} come from
the same Higgs self--energy diagram in fig.~\ref{exsd}.
Now the above cancellation means that the vertex and the fermion propagator
in the boson self--energy of fig.~\ref{exsd} are arranged in such a way that
the last fermion--loop integration is not quadratic divergent.

One could argue that the vertex and the propagators contain different
formfactors and therefore different cutoffs.
As shown in the preliminary approximation and in the full analysis below
the $\zeta$'s and hence the formfactors drop out, at least in the BHL--model.
The reason is that the BHL--model is dominated by the single scale $\Lambd$.
A special top condensation model needs a more precise analysis to proof the
formfactor independence.

On the contrary top condensation generates the \sm as an effective theory.
How can the absence of quadratic divergences be translated into the picture
of the effective theory? Does the above cancellation mean that the Veltman
condition in the \sm is fulfilled and that $M_H\approx 2m_t$?
The effective theory contains a fundamental four--boson--vertex. Hence one
has to use different cutoffs for fermion and boson--loops since the Veltman
condition is regularization scheme dependent.
We get in lowest order:
\bea  4m_t^2\La_t^2-M_H^2\La_H^2 \approx 0  \eea
or
\bea  \frac{\La_t}{\La_H} \approx \frac{M_H}{2m_t}. \eea
The knowledge of the cancellation of quadratic divergences in the low energy
limit does not help to fix the masses in the effective theory and only
predicts the ratios of the different cutoffs. This is precisely the difference
between the cancellation of quadratic divergences in a theory, where the new
interaction is specified, and the Standard Model, where it is not.
That difference is implied by the cutoff notation $\Lambd$ and $\La$.


\section{Cancellation to all orders}

The aim of this section is the step by step development of the quadratic
divergences of the full set of diagrams in fig.~\ref{aoqd}, i.~e. we have to
determine the full four--boson--vertex. It is the solution of a complicated
system of Schwinger--Dyson--equations. For the quadratic divergences it is
sufficient to consider a subsystem of diagrams, which will be established
and solved in the following.

Before we discuss the full four--boson--vertex we have to consider higher 
order contributions to the kinetic part of the boson self--energy. In other 
words: does the $\zeta$ get higher order corrections?
We have to look for $p^2\ln\frac{\Lambd^2}{p^2}$--terms in the boson
self--energy of fig.\ref{aoqd}. As we will see in this section
the full four--boson--vertex is proportional to $\ln\frac{\Lambd^2}{p^2}$, so
that the second graph cannot provide $p^2\ln\frac{\Lambd^2}{p^2}$--terms.
Thus the lowest order kinetic term of the first graph and therefore $\zeta$
does not get further corrections.

Now the full four--boson--vertex contains graphs which consist of boson
propagators and top--loop induced boson--vertices. It seems to be
a Sisyphus work to find all four--boson--diagrams which contribute to the
leading divergence. In the \sm this would be surely not possible.
In top condensation, however, the boson propagators, which contain a factor
$\zeta^{-1}$, partly remove logarithmic divergences so that
$\(\ln\frac{\Lambd^2}{p^2}\)^n$--terms with $n\ge 2$ never appear.
As a consequence we only have to extract simple logarithmic divergent 
diagrams.

The best way to see which graphs are involved is to use Euler's polyhedron
theorem:
\bea  \mbox{\it corners }+\mbox{\it plains }=\mbox{\it edges }+2  \eea
which give a relation for the number of {\it corners, plains} and
{\it edges} of a polyhedron. In our diagram the {\it corners} are the
boson {\it vertices}, the {\it edges} are the boson {\it propagators} and
the {\it plains} are the $\mbox{\it loops}+1$ to close the polyhedron.
One finds
\bea  \mbox{\it vertices}+\mbox{\it loops}-\mbox{\it propagators}=1.
      \label{euler} \eea
This is a power equation for $\ln\frac{\Lambd^2}{p^2}$, because the
vertices contain a $\ln\frac{\Lambd^2}{p^2}$, the propagators a
$\(\ln\frac{\Lambd^2}{p^2}\)^{-1}$ and the whole diagram must be
proportional to $\ln\frac{\Lambd^2}{p^2}$.
As a consequence each boson--loop must be logarithmic divergent and may
only consist of two boson propagators.
Hence $\mbox{\it propagators}=2\cdot\mbox{\it loops}$ and together with
eq.~(\ref{euler}) we get $ 4\cdot \mbox{\it vertices}-4 = 2\cdot
\mbox{\it propagators}$ which is only true, if there are only
four--boson--vertices.
A possible graph is shown in fig.~\ref{example}.
%
\begin{figure}[htb]
\begin{center}
\begin{picture}(250,90)(0,0)
\DashLine(20,20)(40,30)3             \DashLine(20,60)(40,50)3
\DashCArc(35,40)(11.2,297,63)3       \DashCArc(45,40)(11.2,117,243)3
\DashCArc(42,30)(21,30,97)3          \DashCArc(42,50)(21,263,330)3
\DashCArc(70,35)(11.2,27,153)3       \DashCArc(70,45)(11.2,207,333)3
\DashCArc(110,40)(30,90,270)3        \DashLine(110,10)(140,10)3
\DashLine(130,70)(140,70)3           \DashCArc(140,50)(20,0,90)3
\DashCArc(140,30)(20,270,0)3
\DashCArc(155,40)(11.2,297,63)3      \DashCArc(165,40)(11.2,117,243)3
\DashCArc(162,30)(21,30,97)3         \DashCArc(162,50)(21,263,330)3
\DashLine(180,40)(200,60)3           \DashLine(180,40)(200,20)3
\DashCArc(115,20)(11.2,297,63)3      \DashCArc(125,20)(11.2,117,243)3
\DashCArc(120,65)(11.2,27,153)3      \DashCArc(120,75)(11.2,207,333)3
\DashLine(110,70)(110,50)3           \DashLine(130,70)(130,50)3
\DashCArc(115,50)(5,0,359)3          \DashCArc(125,50)(5,0,359)3
\DashCArc(110,48)(21,300,7)3         \DashCArc(130,48)(21,173,240)3
\GCirc(40,30){1.5} 0                 \GCirc(40,50){1.5} 0
\GCirc(60,40){1.5} 0                 \GCirc(80,40){1.5} 0
\GCirc(160,30){1.5} 0                \GCirc(160,50){1.5} 0
\GCirc(180,40){1.5} 0                \GCirc(110,70){1.5} 0
\GCirc(130,70){1.5} 0                \GCirc(110,50){1.5} 0
\GCirc(120,50){1.5} 0                \GCirc(130,50){1.5} 0
\GCirc(120,30){1.5} 0                \GCirc(120,10){1.5} 0
\end{picture}
\end{center}
\caption{\it Possible contribution to the full vertex.}
\label{example}
\end{figure}
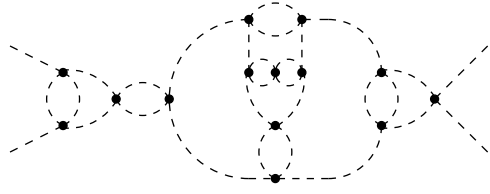
%
\begin{figure}[htb]
\begin{center}
\begin{picture}(460,90)(27,50)
\DashLine(40,60)(80,100)4       \DashLine(40,100)(80,60)4
\GCirc(60,80)4 {0.8}
\Text(31,100)[c]{i}             \Text(31,60)[c]{i}
\Text(90,100)[c]{j}             \Text(90,60)[c]{j}
\Text(110,79)[c]{$=$}
\DashLine(140,60)(180,100)4     \DashLine(140,100)(180,60)4
\GCirc(160,80)2 0
\Text(131,100)[c]{i}            \Text(131,60)[c]{i}
\Text(190,100)[c]{j}            \Text(190,60)[c]{j}
\Text(210,80)[c]{$+$}
\DashLine(240,115)(260,95)4     \DashLine(260,95)(280,115)4
\DashCArc(260,80)(15,90,270)4   \DashCArc(260,80)(15,270,90)4
\DashLine(240,45)(260,65)4      \DashLine(260,65)(280,45)4
\GCirc(260,95)4 {0.2}           \GCirc(260,65)4 {0.2}
\Text(231,115)[c]{i}            \Text(231,45)[c]{i}
\Text(290,115)[c]{j}            \Text(290,45)[c]{j}
\Text(238,80)[c]{i}             \Text(283,80)[c]{j}
\Text(310,80)[c]{$+$}
\DashLine(340,130)(360,110)4    \DashLine(360,110)(380,130)4
\DashCArc(360,95)(15,90,270)4   \DashCArc(360,95)(15,270,90)4
\DashCArc(360,65)(15,90,270)4   \DashCArc(360,65)(15,270,90)4
\DashLine(340,30)(360,50)4      \DashLine(360,50)(380,30)4
\GCirc(360,110)4 {0.2}          \GCirc(360,80)4 {0.2}
\GCirc(360,50)4 {0.2}
\Text(331,130)[c]{i}            \Text(331,30)[c]{i}
\Text(390,130)[c]{j}            \Text(390,30)[c]{j}
\Text(338,95)[c]{i}             \Text(383,95)[c]{j}
\Text(338,65)[c]{i}             \Text(383,65)[c]{j}
\Text(408,80)[l]{$+\;\;\cdots$}
\end{picture}
%
%
\begin{picture}(460,160)(27,10)
\DashLine(40,60)(80,100)4       \DashLine(40,100)(80,60)4
\GCirc(60,80)4 {0.8}
\Text(31,100)[c]{i}             \Text(31,60)[c]{i}
\Text(90,100)[c]{i}             \Text(90,60)[c]{i}
\Text(110,79)[c]{$=$}
\DashLine(140,60)(180,100)4     \DashLine(140,100)(180,60)4
\GCirc(160,80)2 0
\Text(131,100)[c]{i}            \Text(131,60)[c]{i}
\Text(190,100)[c]{i}            \Text(190,60)[c]{i}
\Text(210,80)[c]{$+$}
\DashLine(240,115)(260,95)4     \DashLine(260,95)(280,115)4
\DashCArc(260,80)(15,90,270)4   \DashCArc(260,80)(15,270,90)4
\DashLine(240,45)(260,65)4      \DashLine(260,65)(280,45)4
\GCirc(260,95)4 {0.8}           \GCirc(260,65)4 {0.8}
\Text(231,115)[c]{i}            \Text(231,45)[c]{i}
\Text(290,115)[c]{i}            \Text(290,45)[c]{i}
\Text(238,80)[c]{k}             \Text(283,80)[c]{k}
\Text(310,80)[c]{$+$}
\DashLine(340,130)(360,110)4    \DashLine(360,110)(380,130)4
\DashCArc(360,95)(15,90,270)4   \DashCArc(360,95)(15,270,90)4
\DashCArc(360,65)(15,90,270)4   \DashCArc(360,65)(15,270,90)4
\DashLine(340,30)(360,50)4      \DashLine(360,50)(380,30)4
\GCirc(360,110)4 {0.8}          \GCirc(360,80)4 {0.8}
\GCirc(360,50)4 {0.8}
\Text(331,130)[c]{i}            \Text(331,30)[c]{i}
\Text(390,130)[c]{i}            \Text(390,30)[c]{i}
\Text(338,95)[c]{k}             \Text(383,95)[c]{k}
\Text(338,65)[c]{l}             \Text(383,65)[c]{l}
\Text(408,80)[l]{$+\;\;\cdots$}
\end{picture}
%
%
\begin{picture}(460,80)(27,0)
\DashLine(40,20)(80,60)4        \DashLine(40,60)(80,20)4
\GCirc(60,40)4 {0.2}
\Text(31,60)[c]{i}             \Text(31,20)[c]{i}
\Text(90,60)[c]{j}             \Text(90,20)[c]{j}
\Text(110,39)[c]{$=$}
\DashLine(140,20)(180,60)4      \DashLine(140,60)(180,20)4
\GCirc(160,40)2 0
\Text(131,60)[c]{i}             \Text(131,20)[c]{i}
\Text(190,60)[c]{j}             \Text(190,20)[c]{j}
\Text(200,40)[c]{$+$}
\DashLine(220,20)(240,40)4      \DashLine(240,40)(220,60)4
\DashCArc(255,40)(15,0,180)4    \DashCArc(255,40)(15,180,360)4
\DashLine(290,20)(270,40)4      \DashLine(270,40)(290,60)4
\GCirc(240,40)4 {0.8}           \GCirc(270,40)4 {0.8}
\Text(211,60)[c]{i}             \Text(211,20)[c]{i}
\Text(300,60)[c]{j}             \Text(300,20)[c]{j}
\Text(256,63)[c]{k}             \Text(256,17)[c]{k}
\Text(310,40)[c]{$+$}
\DashLine(330,20)(350,40)4      \DashLine(350,40)(330,60)4
\DashCArc(365,40)(15,0,180)4    \DashCArc(365,40)(15,180,360)4
\DashCArc(395,40)(15,0,180)4    \DashCArc(395,40)(15,180,360)4
\DashLine(410,40)(430,60)4      \DashLine(410,40)(430,20)4
\GCirc(350,40)4 {0.8}           \GCirc(380,40)4 {0.8}
\GCirc(410,40)4 {0.8}
\Text(321,60)[c]{i}             \Text(321,20)[c]{i}
\Text(440,60)[c]{j}             \Text(440,20)[c]{j}
\Text(366,63)[c]{k}             \Text(366,17)[c]{k}
\Text(396,63)[c]{l}             \Text(396,17)[c]{l}
\Text(448,40)[l]{$+\;\;\cdots$}
\end{picture}
%
%
\begin{picture}(460,80)(27,-10)
\Text(200,30)[c]{$+$}
\DashLine(220,60)(240,40)4      \DashLine(250,15)(290,0)4
\DashCArc(250,30)(14.1,135,250)4
\DashCArc(255,40)(15,0,180)4    \DashCArc(255,40)(15,180,360)4
\DashLine(270,40)(290,60)4      \DashLine(220,0)(260,15)4
\DashCArc(260,30)(14.1,290,45)4
\GCirc(240,40)4 {0.2}           \GCirc(270,40)4 {0.2}
\Text(211,60)[c]{i}             \Text(211,0)[c]{i}
\Text(300,60)[c]{j}             \Text(300,0)[c]{j}
\Text(256,63)[c]{i}             \Text(256,33)[c]{j}
\Text(310,30)[c]{$+$}
\DashLine(330,60)(350,40)4      \DashLine(360,15)(430,0)4
\DashCArc(360,30)(14.1,135,250)4
\DashCArc(365,40)(15,0,180)4    \DashCArc(365,40)(15,180,360)4
\DashCArc(395,40)(15,0,180)4    \DashCArc(395,40)(15,180,360)4
\DashLine(410,40)(430,60)4      \DashLine(330,0)(400,15)4
\DashCArc(400,30)(14.1,290,45)4
\GCirc(350,40)4 {0.2}           \GCirc(380,40)4 {0.2}
\GCirc(410,40)4 {0.2}
\Text(321,60)[c]{i}             \Text(321,0)[c]{i}
\Text(440,60)[c]{j}             \Text(440,0)[c]{j}
\Text(366,63)[c]{i}             \Text(366,33)[c]{j}
\Text(396,63)[c]{i}             \Text(396,33)[c]{j}
\Text(448,30)[l]{$+\;\;\cdots$}
\end{picture}
%
%
\begin{picture}(460,80)(27,0)
\DashLine(40,20)(80,60)4        \DashLine(40,60)(80,20)4
\GCirc(60,40)4 {0.2}
\Text(31,60)[c]{i}             \Text(31,20)[c]{i}
\Text(90,60)[c]{i}             \Text(90,20)[c]{i}
\Text(110,39)[c]{$=$}
\DashLine(140,20)(180,60)4      \DashLine(140,60)(180,20)4
\GCirc(160,40)2 0
\Text(131,60)[c]{i}             \Text(131,20)[c]{i}
\Text(190,60)[c]{i}             \Text(190,20)[c]{i}
\Text(200,40)[c]{$+$}
\DashLine(220,20)(240,40)4      \DashLine(240,40)(220,60)4
\DashCArc(255,40)(15,0,180)4    \DashCArc(255,40)(15,180,360)4
\DashLine(290,20)(270,40)4      \DashLine(270,40)(290,60)4
\GCirc(240,40)4 {0.8}           \GCirc(270,40)4 {0.8}
\Text(211,60)[c]{i}             \Text(211,20)[c]{i}
\Text(300,60)[c]{i}             \Text(300,20)[c]{i}
\Text(256,63)[c]{k}             \Text(256,17)[c]{k}
\Text(310,40)[c]{$+$}
\DashLine(330,20)(350,40)4      \DashLine(350,40)(330,60)4
\DashCArc(365,40)(15,0,180)4    \DashCArc(365,40)(15,180,360)4
\DashCArc(395,40)(15,0,180)4    \DashCArc(395,40)(15,180,360)4
\DashLine(410,40)(430,60)4      \DashLine(410,40)(430,20)4
\GCirc(350,40)4 {0.8}           \GCirc(380,40)4 {0.8}
\GCirc(410,40)4 {0.8}
\Text(321,60)[c]{i}             \Text(321,20)[c]{i}
\Text(440,60)[c]{i}             \Text(440,20)[c]{i}
\Text(366,63)[c]{k}             \Text(366,17)[c]{k}
\Text(396,63)[c]{l}             \Text(396,17)[c]{l}
\Text(448,40)[l]{$+\;\;\cdots$}
\end{picture}
\end{center}
\caption{\it Equation system ($i,j,k,\ldots\in\{H,G^0,G^1,G^2\}$ 
and $i\ne j$).}
\label{eqsy}
\end{figure}
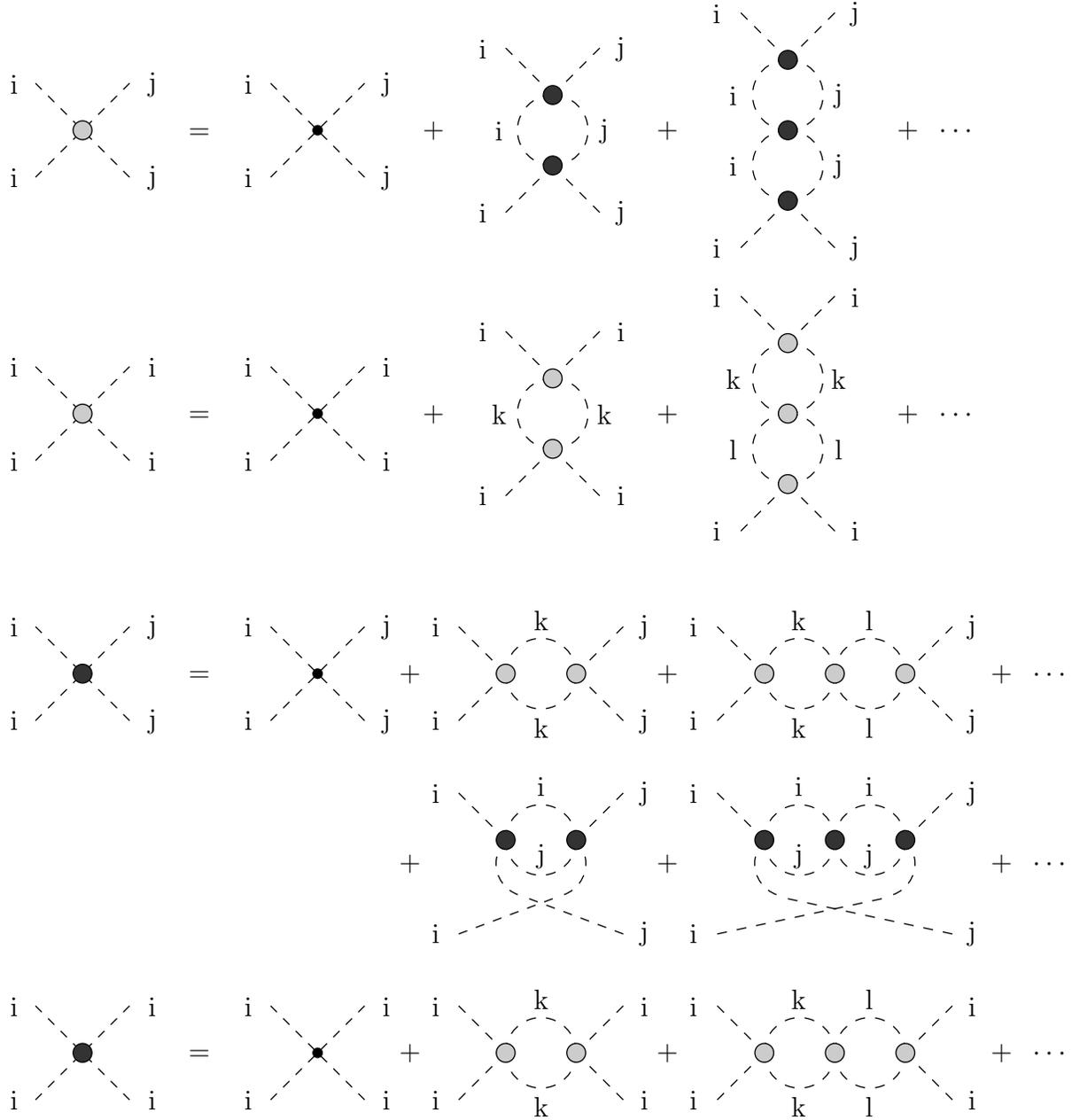
The diagram does not only contain boson--loops with two boson lines.
But if one calculates all such loops with two boson lines, contracts them
to a point and repeats this procedure enough times, one can contract the
whole graph to a point. Therefore two boson lines belong to each loop.
To construct all possible diagrams one has to go just the other way round.
We start with t-- and s--channel bubble chains and replace the vertices
again by bubble sums and so on.
We end up with the self--consistent system of Schwinger--Dyson--equations
in fig.~\ref{eqsy}, where we sum over $k,l,\ldots$. The indices run from
1 to 4, which denote the four Higgs field components $H$, $G^0$, $G^1$, 
$G^2$.

\begin{figure}[htb]
\begin{center}
\begin{picture}(250,90)(0,0)
\DashLine(20,20)(60,60)4             \DashLine(20,60)(60,20)4
\DashLine(90,20)(130,60)4            \DashLine(90,60)(130,20)4
\DashLine(160,20)(200,60)4           \DashLine(160,60)(200,20)4
\DashLine(230,20)(270,60)4           \DashLine(230,60)(270,20)4
\GBox(30,30)(50,50){0.5}
\GCirc(110,40)4 {0.8}
\GCirc(180,40)4 {0.2}
\GCirc(250,40)2 0
\Text(75,40)[c]{$=$}                   \Text(145,40)[c]{$+$}
\Text(216,40)[c]{$-$}
\end{picture}
\end{center}
\caption{\it Equation for the full vertex.}
\label{fullvert}
\end{figure}
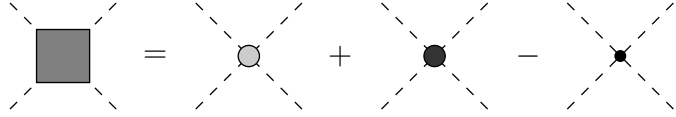

There are three types of vertices denoted by a dot, a bright circle and
a dark circle.
The dot is the usual fermion--loop. One therefore finds:
\bea  \Gamma_{dot}^{ij}=\Gamma_{SM}^{ij}\cdot\zeta\(\frac{2m_t}{M_H}\)^2
      =\(-i2g_t^2\zeta\)\cdot
      A_{ij}\;\;\;\;\mbox{where}\;\;\; A=\(\begin{array}{cccc}
      3 & 1 & 1 & 1 \\ 1 & 3 & 1 & 1 \\ 1 & 1 & 3 & 1 \\ 1 & 1 & 1 & 3
      \end{array}\) \label{ag}  \eea
and $i,j=1,\ldots,4$ run over the Higgs field components\footnote{E.g.
$\Gamma_{\rm SM}^{12}$ is the four--boson--vertex of the \sm with two 
Higgs bosons $H$ and two Goldstone bosons $G^0$.}.
The matrix $A$ contains the usual combinatorical factors. The bright and 
dark circles are auxiliary vertices which are determined by
the equation system. As the dot one can attach a matrix to each of them:
\bea  \Gamma_{bright}^{ij}
      = \(-i2g_t^2\zeta\)\cdot K_{ij} \;\;\;\;\;\;
      \Gamma_{dark}^{ij}
      = \(-i2g_t^2\zeta\)\cdot Q_{ij}.
\eea
The full vertex
\bea  \Gamma_{full}^{ij}
      &=& \(-i2g_t^2\zeta\)\cdot \Gamma_{ij}
\eea
contains s-- and t--channel bubble sums and is therefore
the sum of the bright and the dark vertex minus the double counted
dot (fig.~\ref{fullvert})\footnote{The t--channel bubble sum of the bright
vertex inserted in the second diagram of fig.~\ref{aoqd} provides graphs,
which can be interpreted as boson self--energy corrections for the one--loop
diagram. One could argue that this leads to a double counting of diagrams.
But since the quadratic divergences of the whole diagram and the boson
self--energy subdiagram have different origins there is not any double 
counting of quadratic divergences because we extract different terms.}.
%

Each loop integral $I$ in fig.~\ref{eqsy} is calculated without 
combinatorical factors in the following way:
\bea I=\int\frac{d^4k}{(2\pi)^4}\(-i2g_t^2\zeta\)^2\(\zeta^{-1}\frac{i}{k^2}\)
     \(\zeta^{-1}\frac{i}{(q-k)^2}\). \eea
The leading divergence is extracted for large loop momentum $k$. Thus
all $\zeta$'s in the vertices and the propagators depend on the largest 
momentum $k$ and cancel therefore in this limit. Hence we get:
\bea I=i\frac{4g_t^4}{(4\pi)^2}\ln\frac{\Lambd^2}{p^2}=
      \(-\frac{2}{N_c}\)\(-i2g_t^2\zeta \)  \eea
up to finite terms, where $p$ is the largest momentum in the diagram\footnote{
This $p$ need not to be $q$ but can also come from the infrared behaviour of a 
$\zeta$.}.
This again has the form of a boson vertex with an additional factor 
$\(-\frac{2}{N_c}\)$. In this way we can calculate a complicated diagram step 
by step.

Including all combinatorical factors the equation system of
fig.~\ref{eqsy} reads now ($i$, $j$ fixed, $i\ne j$):
\bea K_{ii}=Q_{ii}= A_{ii}&-&\frac{2}{N_c}\(K^2\)_{ii}+\frac{2}{N_c^2}
     \nonumber        \(K^3\)_{ii}-\frac{2}{N_c^3}\(K^4\)_{ii}+-\ldots \\
     K_{ij}       = A_{ij}&-&\frac{4}{N_c}\(Q_{ij}\)^2+\frac{8}{N_c^2}
     \label{es}       \(Q_{ij}\)^3-\frac{16}{N_c^3}\(Q_{ij}\)^4+-\ldots \\
     Q_{ij}       = A_{ij}&-&\frac{1}{N_c}\(K^2\)_{ij}+\frac{1}{N_c^2}
     \nonumber        \(K^3\)_{ij}-\frac{1}{N_c^3}\(K^4\)_{ij}+-\ldots \\
                          &-&\frac{2}{N_c}\(Q_{ij}\)^2+\frac{4}{N_c^2}
     \nonumber        \(Q_{ij}\)^3-\frac{8}{N_c^3}\(Q_{ij}\)^4+-\ldots \eea
and the full vertex of fig.~\ref{fullvert} has the form:
\bea \Gamma &=& \frac{1}{2}(K-A)+(Q-A)+A  \label{ve} \eea
where we have to put in the solutions of the equation system\footnote{
In line $1$ and $3$ of the system (\ref{es}) the factor $2$ of the expansion
parameter $-\frac{2}{N_c}$ is missing since each boson--loop gets a
combinatorical factor $1/2$. In addition line 1 and 2 have an overall factor 
2. This considers the fact that a reinsertion of a bubble sum destroys a
combinatorical factor $1/2$ in the underlying bubble sum. That factor $2$
must be removed in the vertex equation (\ref{ve}) since there is no further
insertion in a bubble sum.}.

The equations (\ref{es}) are very complicated. It is reasonable to make an
Ansatz for the matrices $K$ and $Q$. We use the assumption that the
solution is invariant under the exchange of two Higgs field components.
In that case $K$ and $Q$ have the following simple form:
\bea                     K=\(\begin{array}{cccc}
      x & y & y & y \\ y & x & y & y \\ y & y & x & y \\ y & y & y & x
      \end{array}\) \;\;\; , \;\;\;\;     Q=\(\begin{array}{cccc}
      u & v & v & v \\ v & u & v & v \\ v & v & u & v \\ v & v & v & u
      \end{array}\) \; . \eea
The equations (\ref{es}) and (\ref{ve}) can be simplified to:
\bea  3x &=& 3u = 3+2N_c-2N_c\cdot \frac{1+\frac{x}{N_c}+2\frac{y}{N_c}}
      {\(1+\frac{x}{N_c}\)^2+2\(1+\frac{x}{N_c}\)\frac{y}{N_c}
      -3\(\frac{y}{N_c}\)^2}  \nonumber \\  \label{xvy}
      y  &=& 1-\frac{\frac{4}{N_c}v^2}{1+\frac{2}{N_c}v} \\
      2v &=& 1-y+\frac{2y}
      {\(1+\frac{x}{N_c}\)^2+2\(1+\frac{x}{N_c}\)\frac{y}{N_c}
      -3\(\frac{y}{N_c}\)^2}  \nonumber \eea
and
\bea  \Gamma_{ii}=3\(\frac{x}{2}-\frac{1}{2}\)   \;\;\;\;\;\;\;\;\;\;
      \Gamma_{ij}=\frac{y}{2}+v-\frac{1}{2}.     \eea
The best way to solve this system is to subtract the third from the
first equation in (\ref{xvy}). One gets:
\bea  (x-y)^2-(x-y)\frac{2}{3}\(v-y+1-\frac{N_c}{2}\)-\frac{2}{3}
      N_c(v-y+1)=0 \eea
and
\bea  (x-y)^2-(x-y)\frac{\frac{4}{N_c}v^2-\frac{N_c}{3}}{1+\frac{2}{N_c}v}
      -N_c\frac{\frac{4}{N_c}v^2+\frac{2}{3}v}{1+\frac{2}{N_c}v}=0 \eea
after replacing of $y$ by the second equation in (\ref{xvy}).
This equation has two types of solutions:
\bea  x-y &=& 2v \label{O4} \\
      x-y &=& -\frac{\frac{N_c}{3}+2v}{1+\frac{2}{N_c}v}. \label{O4b} \eea
The solution (\ref{O4}) is significant for the full vertex $\Gamma$. It gives
$\Gamma_{ii}=3\cdot \Gamma_{ij}$ so that
\bea    \Gamma = \frac{x-1}{2}A,  \label{GA} \eea
i.~e. $\Gamma$ is proportional to the tree level vertex $\Gamma_{SM}$
(see eq.~(\ref{ag})),
which guarantees the O(4)--symmetry\footnote{The appearance of the
O(4)--symmetry in the solutions also confirms the correctness of the
Schwinger--Dyson--system in fig.~\ref{eqsy}.} of the full vertex $\Gamma$.
The solution (\ref{O4b}) does not work in the same way and breaks the
O(4)--symmetry explicitly, which we will not further consider.

The equation system (\ref{xvy}) together with eq.~(\ref{O4}) leads to:
\bea 6x^3-2(N_c+6)x^2-(N_c^2+4N_c-6)x+3N_c^2+6N_c=0. \label{x} \eea
For the physical relevant case $N_c=3$ we have the three solutions:
\bea  x_1=3 \;\;\;\;\; x_2=\sqrt{\frac{5}{2}} \;\;\;\;\;
      x_3=-\sqrt{\frac{5}{2}}.  \label{x123} \eea
After insertion of the first solution in eq.~(\ref{GA}) one gets:
\bea   \Gamma = A   \eea
which is the desired result that the full vertex is just the fermion--loop
induced vertex.
This means that the graphs in fig.~{\ref{aoqd}} and fig.~{\ref{prap}} give
the same result for the leading divergences. The quadratic divergences
cancel therefore automatically as in the preliminary approximation.
The other solutions in eq.~(\ref{x123}) are energetically suppressed since 
they cannot lower the top mass many orders of magnitude.

It is a quite amazing result after the penetration across a technical jungle
and one has to ask for a simple understanding.
In such a case one usually presumes a hidden symmetry responsible for the
cancellation. In fact it is more complicated than expected because only one
solution may have this symmetry while the other ones provide quadratic
divergences. Moreover the x--value of this solution and therefore the boson
loop diagram of fig.~\ref{aoqd} grows proportional to $N_c$ which contradicts
a simple counting of powers of $N_c$. Thus the solution is non--perturbative in
$1/N_c$. A cancellation order by order is therefore excluded and the counting
of degrees of freedom is non--trivial. The following generalization of the
model sheds some light on the subject.


\section{General cancellation condition}

One can ask whether a cancellation occurs for more general gauge group
combinations. For a general framework one can also
generalize the weak gauge group $SU(2)_L$ so that we get a general Standard
Model:
\bea   SU(N_c)_c\times SU(N_L)_L\times U(1)_Y. \label{gega}  \eea
This means that $\psi_L$ in the top condensation Lagrangian (\ref{l}) has
$N_L$ components and therefore $N_L-1$ left--handed b--quarks:
\bea  \psi_L=\(\begin{array}{c} t_L \\ b_{L,1} \\ \vdots \\ b_{L,N_L-1}
      \end{array}\). \label{psi} \eea
The $SU(N_L)_L$ is broken at a high scale in such a way that each left--handed
b--quark can only couple to the $t_L$ at the electroweak scale, so that we 
have $2N_L-1$ massless vector--bosons.
In other words: we have $N_L-1$ $SU(2)_L$--subgroups with one Z--boson and
$2N_L-2$ $W$--type vector--bosons.
They acquire a mass by the top condensation
mechanism, which contains now $2N_L$ Higgs field
components $H,G^0,G^+_1,G^-_1,\ldots,G^+_{N_L-1},G^-_{N_L-1}$,
which are the boundstates $\bar{t}t$, $\bar{t}\gamma^5 t$,
$\bar{b}_{L,1} t_R$, $\bar{t}_R b_{L,1}$, $\ldots$ ,
$\bar{b}_{L,N_L-1} t_R$, $\bar{t}_R b_{L,N_L-1}$,
so that $A$, $K$ and $Q$ are $2N_L$$\times$$2N_L$--matrices:
\bea  A=\(\begin{array}{cccc}
      3 & 1 & \cdots & 1 \\ 1 & \ddots & \ddots & \vdots \\
      \vdots & \ddots & \ddots & 1 \\ 1 & \ldots & 1 & 3
      \end{array}\) \;\;\;\;\;
      K=\(\begin{array}{cccc}
      x & y & \cdots & y \\ y & \ddots & \ddots & \vdots \\
      \vdots & \ddots & \ddots & y \\ y & \ldots & y & x
      \end{array}\) \;\;\;\;\;
      Q=\(\begin{array}{cccc}
      u & v & \cdots & v \\ v & \ddots & \ddots & \vdots \\
      \vdots & \ddots & \ddots & v \\ v & \ldots & v & u
      \end{array}\).   \eea
Hence the system (\ref{xvy}) has the form:
\bea  3x &=& 3u = 3+2N_c-2N_c\cdot \frac{1+\frac{x}{N_c}+(2N_L-2)\frac{y}{N_c}}
      {\(1+\frac{x}{N_c}\)^2+(2N_L-2)\(1+\frac{x}{N_c}\)\frac{y}{N_c}
      -(2N_L-1)\(\frac{y}{N_c}\)^2}  \nonumber \\
      y  &=& 1-\frac{\frac{4}{N_c}v^2}{1+\frac{2}{N_c}v} \\
      2v &=& 1-y+\frac{2y}
      {\(1+\frac{x}{N_c}\)^2+(2N_L-2)\(1+\frac{x}{N_c}\)\frac{y}{N_c}
      -(2N_L-1)\(\frac{y}{N_c}\)^2}.  \nonumber \eea
The general equation for eq.~(\ref{x}) is
\begin{equation}
     2(N_L+1)x^3-2(N_c+2N_L+2)x^2+(2-2N_L(N_c-1)-N_c^2)x+2N_c(N_L+1)+3N_c^2
     =0\label{g1}
\end{equation}
and the general cancellation condition reads
\bea  \frac{x-1}{2} = \frac{N_c}{N_L+1}.  \label{g2} \eea
After the elimination of $x$ in eq.~(\ref{g1}) and (\ref{g2}) we find the
simple condition:
\bea   N_c = \(N_L+1\)\frac{N_L-1}{3-N_L}.  \label{gc}  \eea
One easily sees that the only reasonable and possible combination is:
\bea    N_c=3 \;\;\; , \;\;\;\;\; N_L=2.  \eea
We recognize that the \sm is the only possible choice of gauge groups leading
to a cancellation of quadratic divergences in its top condensation
extension.

In eq.~(\ref{gc}) a special feature of our solution is visible: For 
$N_L\!=\!3$ $N_c$ will be infinite. One would expect that in this case the 
lowest order $1/N_c$ yields the exact results. But there is no cancellation 
of quadratic divergences in lowest order since no boson--loop graphs appear.
The reason for the cancellation is that the above solution is non--perturbative
in $1/N_c$, i.~e. even, if $N_c$ goes to infinity, higher order contributions
are necessary to include.

From eq.~(\ref{gc}) we further see that the reason for the cancellation
is more complicated than counting degrees of freedom as in supersymmetry.
The condition (\ref{gc}) rather reminds one of the anomaly cancellation.
One can try to find a general anomaly condition for our model. Since 
$SU(N_L)_L$ is broken and each $b_{i,L}$ couples only to the $t_L$ one can 
simply associate the top with the charge $2/3$ and all $b$'s with the charge 
$-1/3$. In the same way we have one chargeless tau--neutrino and $N_L-1$ taus 
with charge $-1$.
Thus the anomaly cancellation reads:
\bea  N_c\Bigg[\frac{2}{3}-\frac{1}{3}\cdot\(N_L-1\)\Bigg]+
         \Bigg[0-1\cdot\(N_L-1\)\Bigg]=0 \eea
or
\bea N_c=3\cdot\frac{N_L-1}{3-N_L}.  \label{ac} \eea
Although there is no obvious connection, the conditions (\ref{gc}) and
(\ref{ac}) have a surprising similarity.


\section{Vector--boson contributions}

\begin{figure}[htb]
\begin{center}
\begin{picture}(250,110)(70,-10)
\DashArrowLine(125,20)(155,20)4      \DashArrowLine(185,20)(215,20)4
\PhotonArc(170,55)(25,280,240)3 8
\GCirc(170,27){15} 1
\ArrowArc(170,27)(15,45,135)         \ArrowArc(170,27)(15,135,225)
\ArrowArc(170,27)(15,225,315)        \ArrowArc(170,27)(15,315,45)
\Text(120,20)[c]{$H$}                \Text(220,20)[c]{$H$}
\Text(200,55)[l]{$Z^0,W^\pm$}
\end{picture}
\end{center}
\caption{\it A possible vector--boson contribution.}
\label{vector}
\end{figure}
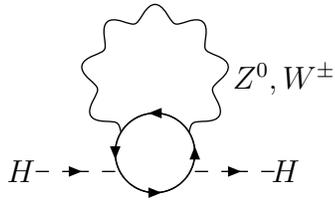

In fig.~\ref{prap} we have neglected the vector--boson contributions.
In the \sm in lowest order of the gauge couplings one has to add the $Z^0$-- 
and $W^\pm$--induced Higgs self--energies in fig.~\ref{qudi}, where in 
lowest order top condensation the four--boson--vertex again
must be replaced by a fermion--loop (see fig.~\ref{vector}).
%
%
These diagrams seem to destroy our cancellation mechanism since they provide
additional quadratic divergences with a negative sign. But these diagrams
are suppressed at least by a factor $M_W^2/m_t^2$ for each vector--boson
line. One can therefore use perturbation theory to extract the vector
boson contributions.
On the contrary both graphs in fig.~\ref{prap} are of the same order in
top condensation and the vector--boson graph in fig.~\ref{vector}
can be regarded as a quantum correction to the first diagram in 
fig.~\ref{prap}. Thus a quantum correction to the second diagram in 
fig.~\ref{prap} is of the same order as the graph in fig.~\ref{vector} but has 
the opposite sign. We get therefore vector contributions with both signs.
Most of these corrections are self--energies or vertex corrections, which 
enter the top mass and the top Yukawa coupling. Hence they drop out for the 
quadratic divergences.

Although one cannot prove the cancellation of the quadratic divergences
of the vector--boson graphs
diagrammatically to all orders, the above consideration shows that these
diagrams do not directly provide quadratic divergences as in the Standard
Model. They can only appear as quantum corrections.
But even if the cancellation mechanism is broken by radiative corrections,
a smaller hierarchy structure remains, i.~e. nevertheless the top condensation
scale could be much larger than 1 TeV.


\section{Conclusion}

We have discussed the hierarchy problem and the corresponding quadratic 
divergences in top condensation models. As we know from supersymmetry a 
cancellation of these quadratic divergences needs certain relations for 
the couplings and the degrees of freedom between the fermions and the 
bosons. In supersymmetry this is guaranteed by symmetry.
Here compositeness leads to a relation between the
top Yukawa--coupling and the Higgs self--coupling in top condensation. 
In analogy to the Veltman condition, which has a similar structure, a
cancellation of quadratic divergences should appear, at least partly.
Nambu called such a scenario ``Quasi--Supersymmetry'', because of the partly
cancellation of the top and the Higgs loop diagrams, but here we could not
identify a symmetry as the origin of the cancellation. The exact cancellation
is not obvious and needs a precise consideration of the contributing
diagrams. The $1/N_c$--expansion does not respect the
Veltman--like cancellation since fermion and boson--loop diagrams are of
different order $1/N_c$. Moreover the boson--loop diagrams are hidden in the
full Schwinger--Dyson--equation and must be extracted by reshuffling of the
Feynman diagrams.
In a preliminary approximation we found that the fermion and boson--loop
graphs in fig.~\ref{prap} are free of quadratic divergences. This 
cancellation does however not lead to a Veltman--like relation between the 
top and the Higgs mass.

To demonstrate the cancellation to all orders we have
shown that the full four--boson--vertex is just the fermion--loop so
that the leading quadratic divergences of fig.~\ref{aoqd} cancel in the same
way as in fig.~\ref{prap}. The full vertex is a solution of a complicated
system of Schwinger--Dyson--equations. It contains indeed a solution where a
miraculous cancellation happens. The fact that it only works for $N_c=3$
and $N_L=2$ means that a possible Grand Unified gauge group at the
Grand Unification scale prefers the Standard Model gauge
group, since the masses of the \sm can be many orders of magnitude smaller
than in other breaking schemes.
If a symmetry could be found being responsible for the cancellation 
mechanism, it would be non--trivial and would have a complicated structure
since it works only for a special solution of the 
Schwinger--Dyson--equations.
The cancellation rather seems to come from a counting of fermionic and
bosonic degrees of freedom depending on the color, flavor and other
statistical factors.
Although there is no better understanding for this mechanism, we 
made some consistency checks for the four--boson--vertex and the
Schwinger--Dyson--equations and the cancellation seems to work.

In this paper we have concentrated our analysis on top condensation models,
which have the four--fermion--interaction (\ref{l}) in the low energy limit,
especially the topcolor--model by C.~T.~Hill. Other models use a different 
color structure, where color is summed between equal quark flavors.
The quadratic divergences of such models can in principle be calculated.
It requires an additional analysis of the color structure in each Feynman
diagram, but a similar cancellation of quadratic divergences is not excluded.

This cancellation mechanism is further not restricted on top condensation.
It can probably be applied to other composite Higgs models. A
cancellation of quadratic divergences in the low energy limit of some
gauge theories in the technicolor scenario would admit a technicolor
scale above the TeV--range.
But especially the naturalness of top condensation would show that a dynamical
origin of the electroweak symmetry breaking enables a simple explanation
of the mass hierarchies.

\vspace*{2cm}

{\bf Acknowledgement:} I would like to thank M.~Beneke,  
M.~Hutter, M.~Misiak and B.~Stech for useful discussions and C.~T.~Hill and 
M.~Lindner for useful comments on the draft version of this paper.



\end{document}